\documentclass[aip,cha,twocolumn,reprint, english]{revtex4-1}
\usepackage{babel}

\usepackage{lmodern}
\usepackage{microtype}

\usepackage{amssymb, amsmath}
\usepackage{bm}
\usepackage{graphicx}

\newcommand{\ignore}[1]{}
\renewcommand{\vec}[1]{\bm{#1}}
\newcommand{\mat}[1]{\bm{#1}}
\newcommand{\domain}{\ensuremath{\Omega}}
\newcommand{\map}{\ensuremath{\vec{F}}}
\newcommand{\koopman}{\mathcal{K}}

\newcommand{\expect}{\mathbb{E}}

\newcommand{\setX}{f_X}
\newcommand{\setY}{f_Y}
\newcommand{\sech}{\operatorname{sech}}
\newcommand{\domainX}{\Omega_X}
\newcommand{\domainY}{\Omega_Y}

\newcommand{\cutoffX}{\varepsilon_X}
\newcommand{\cutoffY}{\varepsilon_Y}

\makeatother

\begin{document}

\title{Identifying Finite-Time Coherent Sets from Limited Quantities
  of Lagrangian Data}

\author{Matthew O. Williams}
\affiliation{Program in Applied and Computational Mathematics, Princeton
  University, NJ 08544.}
\author{Irina I. Rypina}
\affiliation{ Department of Physical Oceanography, Woods Hole
  Oceanographic Institute,  MA 02543.}
\author{Clarence W. Rowley}
\affiliation{Department of Mechanical and Aerospace Engineering,
Princeton University, NJ 08544.}

\begin{abstract}
A data-driven procedure for identifying the dominant transport
barriers in a time-varying flow from limited quantities of Lagrangian
data is presented.
Our approach partitions state space
into coherent pairs, which are sets of initial conditions chosen to
minimize the number of trajectories that ``leak'' from one set to the 
other under the influence of a stochastic flow field during a
pre-specified interval in time.
In practice, this partition is computed by solving an optimization
problem to obtain a pair of functions whose signs determine set membership.
From prior experience with synthetic, ``data rich'' test problems and
conceptually related methods based on approximations of the
Perron-Frobenius operator, we observe that the functions of  interest typically
appear to be smooth.
We exploit this property by using the basis sets associated with spectral or ``mesh-free'' methods, and as a result, our approach has the potential to more accurately approximate these functions given a fixed amount of data.
In practice, this could enable better approximations of the coherent pairs in problems with relatively limited
quantities of Lagrangian data, which is usually the case with experimental geophysical data.
We apply this method to three examples of increasing complexity: the
first is the double gyre, the second is the Bickley Jet, and the third
is data from numerically simulated drifters in the Sulu Sea.
\end{abstract}

\maketitle

\begin{quotation}
Transport barriers separate a fluid flow into regions with qualitatively
different Lagrangian behaviors, and are important for understanding transport and
stirring processes in geophysical flows.
We present a method for identifying these barriers
by partitioning the state space of the system into coherent
sets that are chosen to minimize the number of trajectories that
``switch sets'' in a given time interval.
There are many conceptual similarities between our approach and
probabilistic methods, but our approach is tailored to problems with
limited quantities of Lagrangian data, which is often the case when
the data come from real instruments such as ``drifters'' released into
the ocean.
In particular, we exploit the apparent smoothness of the functions of interest by employing basis functions associated with spectral or ``mesh-free'' methods, which can converge more rapidly than indicator functions in this regime.
As a result, useful (although not fully converged) approximations of
the coherent sets can be obtained from fewer Lagrangian trajectories
compared to other methods.
This approach is applied to identify coherent sets in three fluid
flows: the double gyre, which is commonly used as a benchmark for
different methods, the Bickley Jet, which is an idealized model for
stratospheric flow, and the third is a realistic numerically generated
near-surface flow in the Sulu Sea.
\end{quotation}

\section{Introduction}

\label{sec:intro}

The identification of transport barriers is an important step in
understanding fluid flows that have complex and often chaotic dynamics.
The locations (or absence) of these barriers helps to determine the mixing
properties of the underlying
flow~\cite{Ottino1990,Aref2002,Wiggins2005}, and has practical implications in a number of engineering contexts
including chemical reactors and combustion~\cite{Ottino1990} as well as ecological applications, such as predicting the extent of oil
spills~\cite{Mezic2010,Olascoaga2012}.
As a result, a number of effective yet conceptually different approaches for extracting these structures have been developed.
Geometric methods focus on the identification of invariant manifolds
and finite-time hyperbolic material lines~\cite{Ma2014,Haller2000a},
and include methods based on Finite Time or Finite Scale Lyapunov Exponents~\cite{Brunton2010,Shadden2005,Rypina2010}
and the associated Lagrangian Coherent Structures~\cite{Haller2000,Rypina2011,Haller2011}, and are perhaps
the most widely used set of approaches at the current time. 
However, there are alternative techniques including variational methods~\cite{Beron2013}, ergodic quotient partitions~\cite{Budivsic2012a}, trajectory complexity measures~\cite{Rypina2011} and Lagrangian descriptors~\cite{Mendoza2014}.

In recent years, probabilistic methods, which use a different
definition of coherence, have proved to be a useful alternative to geometric methods~\cite{Froyland2007,Froyland2010,Froyland2010a,Froyland2013,Bollt2013},
and though they have been applied to general flows, have the advantage of
identifying minimally dispersive regions if the flow happens to be  autonomous or time-periodic~\cite{Froyland2009}. 
Many of these approaches define coherent sets based on the
spectral properties of the Perron-Frobenius operator, which is also referred to as the
transfer
operator~\cite{Froyland2007,Froyland2010,Froyland2010a,Froyland2013,Bollt2013}.
In practice, this information is often obtained by constructing a finite-dimensional
approximation of this operator using the Ulam-Galerkin
method~\cite{Bollt2013,Dellnitz2001,Froyland2007,Froyland2010},
which has been implemented efficiently in software packages such as
GAIO~\cite{Dellnitz2001}. 

Many of these methods also assume that the velocity field that defines the motion of fluid parcels or drifters is available.
In problems where this field is unknown, it can often be
estimated from data using tools such as optical flow~\cite{Horn1981,Mussa-Ivaldi1992}. 
Although these approaches could, in principle, be applied
directly to Lagrangian data, the amount of data required
for an accurate approximation is often too large to be practical in an
experimental setting.
Our ambition in this manuscript is to demonstrate that effective approximations
of coherent sets can be obtained with limited quantities of 
Lagrangian data, and is therefore well suited to experimentally obtained data sets.

We define coherent pairs as the solution to an optimization problem that can be solved numerically using the Singular Value Decomposition (SVD).
The result is a pair of functions whose signs can be used to partition
the data into two sets that minimize the number of  elements that
``leak out'' in a given time interval.
In the limit of infinite data, this problem can be succinctly
expressed as an inner product involving the Koopman
operator~\cite{Koopman1931,Koopman1932,Mezic2005,Budisic2012}, and is 
conceptually similar to the analytical definition presented by \citet{Froyland2013}.
The method here could be thought of as a
different finite-dimensional approximation of this overarching
problem, and similar to the algorithms implemented
in GAIO that approximate the Perron-Frobenius operator using indicator
functions~\cite{Dellnitz2001} or the spectral-collocation method presented in Refs.~\citenum{froyland2013estimating,koltai2011efficient}.

Our approach allows for more freedom in the choice of basis functions, and is compatible with basis sets comprised of indicator functions, (piecewise) polynomials~\cite{Williams2014}, or ``mesh-free'' radial basis functions. 
Although any of these choices could produce useful results, there are some advantages to choosing basis sets other than indicator functions.
Intuitively, this results in the same choice that arises when deciding between a spectral method~\cite{Trefethen2000,Boyd2013}, which typically approximates a linear operator using a set of globally supported set of basis functions, and a finite-volume method~\cite{Leveque2002}, which uses compactly supported functions instead.
Although either choice can produce accurate results, spectral methods typically converge more rapidly than finite-volume methods provided that the functions of interest are smooth~\cite{Trefethen2000,Boyd2013,Leveque2002,Wendland1999}, and from ``data rich'' examples and pre-existing efforts using transfer operator methods~\cite{Bollt2013,Ma2013,Froyland2009,Froyland2010}, this appears to be the case for the functions that define coherent pairs in the applications of interest to us.
The practical benefit of a higher convergence rate is that effective approximations of coherent pairs can be obtained with fewer basis functions, and hence, fewer data points, which implies our approach is well suited for the ``data poor'' regime that often occurs experimentally.

The remainder of the manuscript is outlined as follows: in
Sec.~\ref{sec:conceptual-pair} we give a definition of a coherent set
in terms of a solution to a data-driven optimization problem.
In Sec.~\ref{sec:Koopman}, we consider the infinite data limit, where
this problem can be recast into one involving the Koopman operator.
As a result, methods like Generalized Laplace Analysis~\cite{Mezic2013,Mezic2005,Budisic2012} or Extended Dynamic
Mode Decomposition (Extended DMD)~\cite{Williams2014}, could be re-tasked to compute coherent sets.
Furthermore, this limit makes the connection between our approach and the
analytical definition presented by~\citet{Froyland2013} more clear.
In Sec.~\ref{sec:examples}, we apply our method to three examples
examples: the double gyre, the Bickley Jet, and numerically simulated
drifters in the Sulu Sea, in order to demonstrate that the approach is
effective in practice.
Finally in Sec.~\ref{sec:conclusions}, we present some brief concluding remarks.

\section{A Definition of A Coherent Pair}
\label{sec:conceptual-pair}

In this section, we construct the optimization problem whose
solution defines our pairs of coherent sets.
As we will demonstrate in Sec.~\ref{sec:Koopman}, this approach is
conceptually equivalent to the one presented analytically by
\citet{Froyland2013} and implemented using
GAIO~\cite{Dellnitz2001}.
As a result, there will be many similarities between what was done in
Refs.~\citenum{Froyland2010,Froyland2010a,Froyland2007} and what we do
here; indeed, the manipulations that follow are motivated by Refs.~\citenum{Froyland2010,Froyland2010a,Froyland2007,Froyland2013}. 
The key difference is that those approaches are tailored to use
pre-specified functions such as indicator functions, while our approach is
compatible with any reasonable basis set whose span contains the constant function.

\subsection{The Intuitive Problem}

We assume we are given a collection of $M$ drifters, whose evolution is completely determined by the
velocity field of some underlying flow, 
that are initially contained in some domain, $\domainX\subseteq\domain\subset\mathbb{R}^N$, at time $n$,
but migrate to another domain, $\domainY\subseteq\domain\subset\mathbb{R}^N$, at time $n+1$.
These pairs of positions are collected into the set $\{(\vec x_m, \vec
y_m)\}_{m=1}^M$ where $\vec x_m$ is the position of the $m$-th drifter
at time $n$, and $\vec y_m$ is the position of that drifter at time $n+1$.
Our objective is to partition these drifters into two sets --
$X_1$ and $X_2$ at time $n$ and $Y_1$ and $Y_2$ at time $n+1$ -- based
on their {\em physical positions} at times $n$ and $n+1$ respectively.
To do this we define a pair of functions, $\setX:\domainX\to\pm 1$ and
$\setY:\domainY\to\pm 1$; at time $n$, the sign of $\setX$ determines
whether a point is in $X_1$ or $X_2$, and at time $n+1$, the sign of
$\setY$ is used to assign the data points to either $Y_1$ or $Y_2$.

As shown in Ref.~\citenum{Froyland2013}, the functions
$\setX$ and $\setY$ will only identify useful coherent sets if the
flow that maps $\vec x_m$ to $\vec y_m$ is stochastic.
When this mapping is deterministic, one can find a pair of functions
such that $g(\setX, \setY) = 1$ for any admissible partition of
$\domainX$ and $\domainY$ simply by choosing $Y_1$ to be the image of the set $X_1$, and defining $\setX$ and $\setY$ appropriately.
To produce a pair of {\em distinguished} coherent sets, some stochasticity is required.
Because we will approximate $\setX$ and $\setY$ using relatively small numbers of basis functions, this required ``noise'' is often created implicitly via our choice of basis functions.
However, to ensure that the underlying system appears to be stochastic, we also add explicit but small perturbations to both $\vec x_m$ and $\vec y_m$.
For the problems we will discuss, the choice of the functions used to approximate $\setX$ and $\setY$ appear to have a larger impact on the resulting coherent sets than the externally added noise, but this may not always be the case if a sufficiently large number of basis functions are used.

Intuitively, one wants to choose a pair of coherent sets in a way that minimizes the ``leakage'' that occurs over a finite interval in time, or equivalently, maximizes the number of points that remain within a one of the two sets.
Because the signs of the functions $\setX$ and $\setY$ determine set membership, this intuitive goal can be achieved by choosing $\setX$ and $\setY$ to maximize:
\begin{equation}
  \label{eq:objective}
  g(\setX, \setY) = \frac{1}{M}\sum_{m=1}^M \setX(\vec x_m)\setY(\vec y_m),
\end{equation}
where $g(\setX, \setY) = 1$ if no drifters switch sets.

Without additional constraints, a global maximum can be obtained
trivially by assigning {\em all} the data points to one set or the other.
To force the algorithm to partition the data into two {\em nonempty sets},
we include another pair of constraints that specify the relative
sizes of $X_1$ and $X_2$ and $Y_1$ and $Y_2$.
In particular, we require that:
\begin{equation}
  \label{eq:split-constraint}
\frac{1}{M}  \sum_{m=1}^M \setX(\vec x_m) = \cutoffX, \qquad 
\frac{1}{M}  \sum_{m=1}^M \setY(\vec y_m) = \cutoffY,
\end{equation}
where $\cutoffX$ and $\cutoffY$ are two constants that determine
the difference in the number of elements in $X_1$ and $X_2$ and $Y_1$ and
$Y_2$ respectively.
To obtain two sets of equal size, we set $\cutoffX=\cutoffY=0$, but it is often advantageous to allow $\cutoffX$ and $\cutoffY$ to vary
as not all systems can be (or should be) decomposed into two sets of equal size.

However, even (\ref{eq:objective}) and (\ref{eq:split-constraint}) together is not sufficient to uniquely define $\setX$ and $\setY$.
Indeed, there are either no feasible solutions  (e.g., $M$ is odd and $\cutoffX=\cutoffY=0$) or many optimal solutions (e.g., choose $\setX(\vec x_m) = \setY(\vec y_m)$ for all $m$) when $M$ is finite. 
As a result, further alterations to this intuitive problem are required if a pair of distinguished coherent sets are to be identified.

\subsection{A Finite Dimensional Approximation}

In particular, we will modify the set of admissible $\setX$ and $\setY$.
First, we relax the constraint that $\setX:\domainX\to\pm 1$ and
$\setY:\domainY\to\pm 1$ and allow $\setX:\domainX\to\mathbb{R}$ and
$\setY:\domainY\to\mathbb{R}$.
Next, we approximate  $\setX$ and $\setY$ with functions that lie in the subspace spanned by two sets of basis functions that we denote as $\{\psi_k\}_{k=1}^{K_X}$ and $\{\tilde\psi_k\}_{k=1}^{K_Y}$ for $\setX$ and $\setY$ respectively.

In the discussion that follows, we assume that the first elements in each set are the relevant constant functions,
$\psi_1(\vec x) = 1$ and $\tilde\psi_1(\vec y) = 1$.
This ordering is helpful because it will create a block structure in our finite dimensional approximation that makes it easy to show that the constant function would be a solution to our relaxed optimization problem if some constraints were relaxed.
Furthermore, provided the constant function lies in the span of the basis sets provided, one can always create such a set by ``rearranging'' the basis functions.
Next, we define the vector-valued functions,
\begin{equation}
\vec\psi_X(\vec x) =
	\begin{bmatrix}
	\psi_1(\vec x) = 1 \\
	\psi_2(\vec x) \\
	\vdots \\
	\psi_{K_X}(\vec x)
	\end{bmatrix},\quad 
\vec\psi_Y(\vec y) =
	\begin{bmatrix}
	\tilde\psi_1(\vec y) = 1 \\
	\tilde\psi_2(\vec y) \\
	\vdots \\
	\tilde\psi_{K_Y}(\vec y)
	\end{bmatrix},
	\label{eq:vector-valued-dictionary}
\end{equation}
which allows our finite-dimensional approximations of $\setX$ and $\setY$ to be written as
\begin{equation}
\setX = \sum_{k=1}^{K_X} a_k\psi_k = \vec\psi_X^T\vec a, \quad 
\setY = \sum_{k=1}^{K_Y} \tilde a_k \tilde\psi_k =
\vec{\psi}_Y^T\vec{\tilde a},
\label{eq:solution-ansatz}
\end{equation}
given two vectors of coefficients $\vec a$ and $\vec{\tilde{a}}$.
Because the maximum values of $|\setX|$ and $|\setY|$ are no longer
bounded, we include two additional constraints:
\begin{equation}
  \label{eq:scale-constraints}
  \frac{1}{M}\sum_{m=1}^M |\setX(\vec x_m)|^2 = \frac{1}{M}\sum_{m=1}^M |\setY(\vec y_m)|^2 = 1,
\end{equation}
to impose an overall scaling on both functions.
In all that follows, we will assume that $\setX$ and $\setY$ are smooth functions, and therefore, can be accurately approximated even if $K_X$ and $K_Y$ are relatively small.

\begin{figure*}[t]
\centering \includegraphics[width=0.8\textwidth]{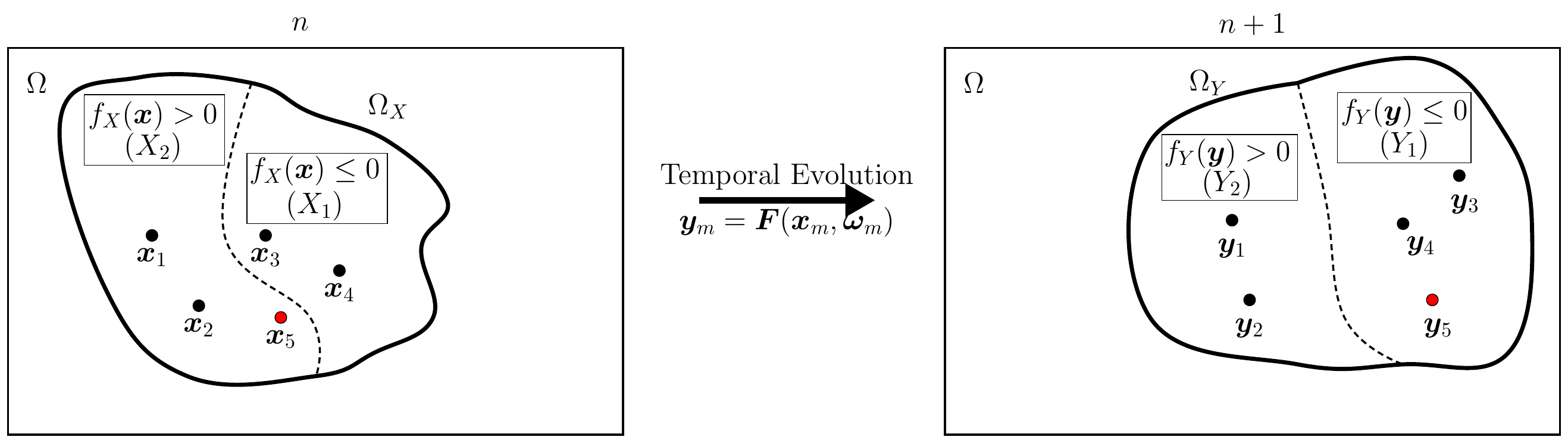}
\caption{A ``cartoon'' of the coherent set definition in this
  manuscript.
Given data from a discrete time dynamical system whose evolution
operator  at time $n$ is $\map$, where $\vec \omega_m$ represents the
``noise'' added to the system,  our objective is
to identify two functions, $\setX:\domainX\to\mathbb{R}$ and
$\setY:\domainY\to\mathbb{R}$, whose signs will be used to partition
$\domainX$ and $\domainY$ into $X_1$ and $X_2$ or $Y_1$ and $Y_2$ respectively. 
These functions are determined by solving
the optimization problem in (\ref{eq:relaxed-discrete}).
Intuitively, the computed functions minimize the number of
``mis-classified'' points, such as $\vec{x}_{5}$, that are assigned to the
two different sets, $X_1$ or $X_2$ and $Y_1$ or $Y_2$, at times $n$ and $n+1$. 
In general, $\domainY\neq\domainX$, so the basis functions
used to approximate $\setX$ and $\setY$ could (and typically should) differ.}
\label{fig:cartoon}
\end{figure*}

With this approximation, the objective function, (\ref{eq:objective}), is:
\begin{equation}
  \label{eq:objective-compressed}
  g(\setX, \setY) = 
   \vec{a}^T \left(\frac{1}{M}\sum_{m=1}^M \vec\psi_X(\vec
     x_m)\vec\psi_Y^T(\vec y_m)\right)\vec{\tilde a} = \vec{a}^T\mat{A}\vec{\tilde{a}}.
\end{equation}
Similarly, the constraints, (\ref{eq:split-constraint}) and  (\ref{eq:scale-constraints}), are:
\begin{subequations}
\label{eq:constraints}
\begin{align}
&   \frac{1}{M}\sum_{m=1}^M \psi_1(\vec x_m)\setX(\vec x_m) =  
   \vec{e}_1^T\mat{G_X}\vec{a} = \cutoffX, \label{eq:constraints:a}\\
&   \frac{1}{M}\sum_{m=1}^M \tilde\psi_1(\vec y_m)\setY(\vec y_m)  = 
   \vec{\tilde{e}}_1^T\mat{G_Y}\vec{\tilde{a}} = \cutoffY, \label{eq:constraints:b}\\
&   \frac{1}{M}\sum_{m=1}^M \setX(\vec x_m)\setX(\vec x_m) = 
%\vec a^T \left(\frac{1}{M} \sum_{m=1}^M\vec\psi(\vec x_m)\vec\psi(\vec x_m)^T\right)\vec{a} =
   \vec{a}^T\mat{G_X}\vec{a} = 1, \\
&   \frac{1}{M}\sum_{m=1}^M \setY(\vec y_m)\setY(\vec y_m) = 
%   \vec{\tilde a}^T \left(\frac{1}{M} \sum_{m=1}^M\vec{\tilde\psi}(\vec y_m)\vec{\tilde\psi}(\vec y_m)^T\right)\vec{\tilde{a}} =
   \vec{\tilde{a}}^T\mat{G_Y}\vec{\tilde{a}} = 1,
\end{align}
\end{subequations}
where $\vec e_1$ and $\vec{\tilde e}_1$ are the first unit vectors in $\mathbb{R}^{K_X}$ and $\mathbb{R}^{K_Y}$ respectively, and 
\begin{subequations}
\label{eq:coherent-matrix}
\begin{align}
\mat{G_X} &\triangleq \frac{1}{M}
\sum_{m=1}^M\vec{\psi}(\vec x_m)\vec{\psi}(\vec x_m)^T,\\
\mat{G_Y} &\triangleq \frac{1}{M} \sum_{m=1}^M\vec{\tilde\psi}(\vec
y_m)\vec{\tilde\psi}(\vec y_m)^T,\\
\mat{A} &\triangleq \frac{1}{M}\sum_{m=1}^M \vec{\psi}(\vec x_m)\vec{\tilde\psi}(\vec y_m)^T.
\end{align}
\end{subequations}
Note that our choice of $\psi_1 = 1$ and $\tilde\psi_1 = 1$ was used in \eqref{eq:constraints:a} and \eqref{eq:constraints:b}.
With this notation, the relaxed,  finite-dimensional optimization problem is:
\begin{subequations} 
\label{eq:relaxed-discrete}
\begin{align}
  \max_{\vec a, \vec{\tilde{a}}} \quad&\vec a^T \mat{A} \vec{\tilde{a}} \\
  \text{subject to:}  \quad    & \vec{e}_1^T\mat{G_X}\vec{a} = \cutoffX, \\
                           &
                           \vec{\tilde{e}}_1^T\mat{G_Y}\vec{\tilde{a}}
                           = \cutoffY, \\
                           & \vec{a}^T\mat{G_X}\vec{a} =
                           \vec{\tilde{a}}\mat{G_Y}\vec{\tilde{a}} = 1.
\end{align}
\end{subequations}
A schematic of \eqref{eq:relaxed-discrete} is given in Fig.~\ref{fig:cartoon}.
In short, the objective is to choose $\setX$ and $\setY$ to maximize the number of data points where $\setX(\vec x_m)$ and $\setY(\vec y_m)$ have the same sign, which is equivalent to minimizing the number of points that switch sets.
The constraints are required to ensure that two \emph{non-empty sets} are identified, and impose an overall scaling on the functions. 
As written, \eqref{eq:relaxed-discrete} is a quadratically-constrained quadratic program, which can be solved~\cite{Albers2011} using specialized numerical routines.
However, we will show that this particular problem can also be solved using the Singular Value Decomposition (SVD).

\subsection{Computing  Coherent Sets in Practice}

In this section, we will show that (\ref{eq:relaxed-discrete}) can be
solved using the SVD.
The motivation for what follows is more mathematical than physical,
and is inspired by the results of \citet{Froyland2013}.
Specifically we note that:
\begin{enumerate}
\item The pair of unit vectors, $\vec u$ and $\vec v$, that maximize the 
quantity $\vec u^T\mat{A}\vec{v}$ are the  left and right singular vectors of
$\mat{A}$ with the largest singular value, which we refer to as
$\vec u_1$ and $\vec v_1$.
\item With the addition of the constraints ${\vec u_1^T \vec u = \cutoffX}$ and
$\vec v_1^T \vec v = \cutoffY$, the optimal solution becomes ${\vec u = \cutoffX \vec u_1 + \sqrt{1 - \cutoffX^2}
\vec u_2}$ and ${\vec v = \cutoffY\vec v_1 + \sqrt{1 - \cutoffY^2}\vec v_2}$, where 
$\vec u_2$ and $\vec v_2$ are the singular  vectors associated with the \emph{second} largest singular value.
\end{enumerate}
The main difference between this problem, which
can be solved using the SVD, and (\ref{eq:relaxed-discrete})
is that the constraints could be written in terms of the standard
Euclidean inner product while (\ref{eq:relaxed-discrete}) has
constraints that are written in terms of weighted inner products.

Therefore, the first step is to transform our coordinates such
that the constraints  in (\ref{eq:relaxed-discrete}) can be expressed
in terms of ``unweighted'' inner products like in our model problem.
To do this, we use the Cholesky Decomposition and let
\begin{equation}
  \label{eq:gram-def}
 \mat{G_X} = \mat{L_X}\mat{L_X}^T, \quad
 \mat{G_Y} = \mat{L_Y}\mat{L_Y}^T.
\end{equation}
For this decomposition to exist,  $\mat{G_X}$ and $\mat{G_Y}$ must be full rank, which will be the case if the sets of functions used to represent $\setX$ and $\setY$ form a basis for a subspace of $L^2(\domainX,\rho)$ and $L^2(\domainY,\nu)$, where $\rho$ and $\nu$ are the spatial distribution of the $\vec x_m$ and $\vec y_m$ respectively.
If we define $\vec b = \mat{L_X}^T\vec a$ and
$\vec{\tilde b} = \mat{L_Y}^T\vec{\tilde a}$, then the constraints simplify to 
\begin{subequations}
\begin{align}
  \vec{e_1}^T\vec{b}&=\hat\cutoffX, \\
  \vec{\tilde e}_1^T\vec{\tilde b}&=\hat\cutoffY, \\
  \vec{b}^T\vec{b} &= \vec{\tilde b}^T\vec{\tilde b} = 1,
\end{align}
\end{subequations}
where $\hat\cutoffX =\cutoffX/\mat{L_X}^{(11)}$, $\hat\cutoffY
=\cutoffY/\mat{L_Y}^{(11)}$, $\mat{L_X}^{(11)}\in\mathbb{R}$ denotes the element
in the first row and column of $\mat{L_X}$.
These terms appear because $\mat{L_X}$ (or $\mat{L_Y}$) is
lower-triangular, and therefore $\vec{e_1}^T\mat{L_X} = \mat{L_X}^{(11)}\vec e_1^T$.
We also rewrite the objective function, and set $\vec
a^T\mat{A}\vec{\tilde{a}} = \vec b^T\mat{\hat{A}}\vec{\tilde b}$ where
\begin{equation}
\label{eq:ahat-def}
\mat{\hat{A}} \triangleq \mat{L_X}^{-1}\mat{A}\mat{L_Y}^{-T}.
\end{equation}
This results in a transformed system of equations
\begin{subequations} 
\label{eq:transformed-relaxed-problem}
	\begin{align}
	\max_{\vec b, \vec{\tilde b}} &\quad
	\vec{b}^T\mat{\hat{A}}\vec{\tilde b},\label{eq:objective-relaxed}\\
	\text{subject to: } 
	& \vec{e_1}^T\vec{b}=\hat\cutoffX,\label{eq:constraint_1} \\
 	& \vec{\tilde e}_1^T\vec{\tilde b}=\hat\cutoffY,\label{eq:constraint_2} \\
 	&\vec{b}^T\vec{b} = \vec{\tilde b}^T\vec{\tilde b} = 1,
	\end{align}
\end{subequations} 
which is formally equivalent to our model problem because $\vec e_1$
and $\vec{\tilde e}_1$ are the left and right singular vectors of
$\mat{\hat{A}}$ with $\sigma_1=1$.
This is a result of our choice of $\psi_1 = \tilde\psi_1 = 1$, and is simple but tedious to show (see Appendix~\ref{app:block-structure}).

Therefore, the solution to (\ref{eq:transformed-relaxed-problem}) is of
the form:
\begin{subequations}
  \label{eq:related-solution}
\begin{align}
\vec b &= \hat{\varepsilon}_X\vec e_1 + \sqrt{1 - \hat\varepsilon_X^2}\vec u_2, \\
\vec{\tilde{b}} &= \hat{\varepsilon}_Y\vec{\tilde{e}}_1 + \sqrt{1 - \hat\varepsilon_Y^2}\vec v_2,
\end{align}
\end{subequations}
where $\vec u_2$ and $\vec v_2$ are the left and right singular
vectors associated with $\sigma_2$, the largest singular value not
equal to $\sigma_1 = 1$.
In what follows, we will assume that $1 = \sigma_1 > \sigma_2$.
  In the un-relaxed problem, the maximum value of the objective function is 1, but in the relaxed problem it is possible to find solutions associated with larger values.
  These solutions are associated with values of $\sigma_2 > 1$, and do not appear to produce useful pairs of sets.
  Instead, we treat the magnitude of the largest singular value as a ``sanity check'' on the procedure.
  If $\sigma_2 > 1$, then our relaxed procedure is identifying solutions that exceed the theoretical maximum of the original problem, and therefore is not a reliable surrogate for the original problem.
  In practice, we have found that reducing the number of basis functions used to approximate both $\setX$ and $\setY$ alleviates this issue.

Once $\vec b$ and $\vec{\tilde{b}}$ have been computed, we let
$\vec a = \mat{L_X}^{-T}\vec b$ and $\vec{\tilde{a}}
=\mat{L_Y}^{-T}\vec{\tilde{b}}$, and approximate $\setX$ and $\setY$ at
any desired points using (\ref{eq:solution-ansatz}).
The final step in the procedure is partition $\domainX$ and $\domainY$
using the numerically computed $\setX$ and $\setY$. 
We define
\begin{subequations}
\begin{align} 
X_{1}&=\{\vec{x}\in\domainX:\setX(\vec{x})\leq0\},\\
Y_{1}&=\{\vec{y}\in\domainY:\setY(\vec{y})\leq0\},
\end{align}
\end{subequations}
and  let $X_{2}$ and $Y_{2}$ be their complements (or, equivalently,
the subset where $\setX,\setY>0$). 
The values of
$\varepsilon_X$ and $\varepsilon_Y$ effectively
add a constant offset to both $\setX$ and $\setY$. 
As a result, $\varepsilon_X$ and $\varepsilon_Y$ 
can  be determined after the fact, and following
Ref.~\citenum{Froyland2010a}, we will choose them so that
the computed $\setX$ and $\setY$ maximize  the fraction of
consistently classified points (i.e., if
$\vec{x}_{m}\in X_{1}$ then $\vec{y}_{m}\in Y_{1}$) in the pairs of
sets ($X_1$, $Y_1$) and ($X_2$, $Y_2$). 
Although we do not place any explicit
constraints on the values of $\cutoffX$ or $\cutoffY$, we typically require 
that neither $\cutoffX$ nor $\cutoffY$ can be so large
or small that either $X_{1}$ or $X_{2}$ (or $Y_{1}$ and $Y_{2}$)
contain a negligible number of data points.

\subsection{Algorithm Summary}
In practice, this algorithm requires the user to provide three
quantities: (i) a data set of snapshot pairs, $\{(\vec x_m, \vec
y_m)\}_{m=1}^M$, (ii) two sets of basis functions that comprise
the vector-valued functions $\vec\psi_X$ and $\vec\psi_Y$, and (iii)
the ``noise'' that will be added to the data. 
The first two quantities are important if this method is to perform
well, but because they are highly problem dependent, we will defer the
discussion of these choices until Sec.~\ref{sec:examples} where we
apply the method to our example problems.
The addition of noise will, in principle, affect the resulting sets,
but in practice, appears to have a smaller impact than the data and
basis functions provided the noise chosen is not too large.
Given these quantities, the coherent sets are computed  as follows:
\begin{enumerate}
\setcounter{enumi}{-1}
\item (Optional) Augment the existing data set with noise by looping
  through the data multiple times and randomly perturbing $\vec x_m$ and $\vec y_m$.
  These new data pairs are then added to the existing set of data, and will be used in the steps that follow.
  In practice, this step is often unnecessary; we effectively inject noise into the problem by using a limited number of basis functions in our approximation of $\setX$ and $\setY$. 

\item Compute the matrices in \eqref{eq:coherent-matrix}, their
  Cholesky decompositions in (\ref{eq:gram-def}), and
  the matrix $\mat{\hat{A}}$ in (\ref{eq:ahat-def}).
\item Using the SVD, let  ${\mat{\hat{A}} =
  \mat{U}\mat{\Sigma}{\mat{V}}^T}$.
\item As a sanity check, examine $\sigma_2$, which is the largest
  singular value that is not unity.
  In practice, we iterate through steps 1-3, and select the largest basis
  where $\sigma_2 < 1$.
\item Choose values of $\cutoffX$ and $\cutoffY$, and compute $\vec b$
  and $\vec{\tilde{b}}$ using (\ref{eq:related-solution}).
  To obtain sets with less ``leakage'', we choose these values to minimize
  the fraction of misclassified points, which is similar to the concept of {\em
    coherence} in Refs.~\citenum{Froyland2010,Froyland2010a}.
  In practice, simply letting $\varepsilon_X=\varepsilon_Y = 0$  is
  sufficient in many applications.
\item Compute $\vec a = \mat{L_X}^{-T}\vec b$ and
  $\vec{\tilde{a}} = \mat{L_Y}^{-T}\vec{\tilde{b}}$, which are the
  solutions of the original relaxed, finite-dimensional optimization problem.
\item Finally, compute the value of $\setX$ or $\setY$ at any desired points using  (\ref{eq:solution-ansatz}),
  and partition the domain based on the sign of $\setX$.
\end{enumerate}
If more than two coherent sets are desired, we repeat the procedure
outlined above in a recursive fashion using the data in $X_1$ and $Y_1$ and the
data in $X_2$  and $Y_2$ separately.
Similar to the work of \citet{Ma2013}, this results in a larger number
of coherent sets that can capture finer spatial features.
In practice, we terminate this iteration procedure if more than 5\% of
the data in any pair of sets leaks out during the interval of
interest so that all of the resulting sets will, visually, appear to
be coherent.

The algorithm presented here runs in $\mathcal{O}(K^2 \max(K,M))$ time, where $K=\max(K_X, K_Y)$.
This cost is either determined by the need to assemble $\mat{G_X}$, $\mat{G_Y}$ and $\mat{A}$, which is an
$\mathcal{O}(K^2M)$ computation, or to decomposing $\mat{\hat{A}}$ which is $\mathcal{O}(K^3)$ operation.
Assuming $K_X \sim K_Y$, this is the same asymptotic complexity as the algorithms used in GAIO if they are naively implemented.
However given equal numbers of basis functions, our approach will be \emph{slower} than GAIO because it uses tree-like data structures to efficiently construct the needed matrices, which our approach is unable to do, and avoids the additional Cholesky factorizations our procedure requires. 
As we shall demonstrate shortly, our approach can often identify useful coherent pairs using far fewer basis functions, which in practice, helps to offset the larger cost-per-basis-function associated with this method.

\section{Connections to the Koopman Operator}
\label{sec:Koopman}
The algorithm presented in Sec.~\ref{sec:conceptual-pair} is both
conceptually and mathematically related to the approach presented in
Refs.~\citenum{Froyland2010,Froyland2010a,Froyland2013}; indeed, the
primary difference between the approaches is that we can use a ``richer''
set of basis functions to represent $\setX$ and $\setY$.
In this section, we examine the ``infinite data'' limit, which is the
limit where our approach can be compared to these transfer
operator-based methods.

In all that follows, we assume our data set $\{(\vec x_m, \vec y_m)\}_{m=1}^M$
is constructed by randomly choosing initial conditions, $\vec x_m$,
from the distribution, $\rho$.
As before, $\vec y_m$ is the location of the $m$-th drifter at time $n+1$, and
$\nu$ is the new distribution at that time.
If the evolution operator from time $n$ to $n+1$ is $\map$, then $\vec
y_m = \map(\vec x_m, \vec\omega_m)$ where $\vec\omega_m$ accounts for both the noise
that is artificially added to the flow map and any stochasticity that naturally exists in the flow.

In the limit as $M\to\infty$, the $ij$-th element of $\mat{G_X}$ is
almost surely:
\begin{equation}
  \label{eq:Gx-limit}
  \lim_{M\to\infty}\mat{G_X}^{(ij)} = \lim_{M\to\infty}
  \frac{1}{M}\sum_{m=1}^M\psi_i(\vec x_m)\psi_j(\vec x_m)
  = \left\langle \psi_i, \psi_j\right\rangle_\rho,
\end{equation}
where $\left\langle f, g \right\rangle_\rho = \int_{\domainX} f(\vec
x) g(\vec x) \rho(\vec x)\;d\vec x$.
This argument follows directly from the law of large numbers.
Similarly, 
\begin{equation}
  \label{eq:Gg-limit}
  \lim_{M\to\infty}\mat{G_Y}^{(ij)} = \lim_{M\to\infty}
  \frac{1}{M}\sum_{m=1}^M\tilde\psi_i(\vec y_m)\tilde\psi_j(\vec y_m) 
  = \left\langle \tilde\psi_i, \tilde\psi_j\right\rangle_\nu,
\end{equation}
where $\left\langle f, g \right\rangle_\nu = \int_{\domainY} f(\vec
y) g(\vec y) \nu(\vec y)\;d\vec y$.
In this limit, both $\mat{G_X}$ and $\mat{G_Y}$ are Gram matrices
where each element is an inner product of basis functions
weighted by the density of the data.

Unlike $\mat{G_X}$ and $\mat{G_Y}$, the $ij$-th element of 
$\mat{A}$ depends upon both the randomly selected initial conditions,
$\vec x_m$, and their images, $\vec y_m$, which are affected by the
stochasticity in the dynamics.
Assuming that the $\vec x_m$ and $\vec\omega_m$ are chosen
independently, 
\begin{align}
  \label{eq:A-limit}
  \lim_{M\to\infty}\mat{A}^{(ij)} &= \lim_{M\to\infty}
  \frac{1}{M}\sum_{m=1}^M\psi_i(\vec x_m)\tilde\psi_j(\vec y_m)\nonumber \\
  &=  \lim_{M\to\infty}\frac{1}{M}\sum_{m=1}^M\psi_i(\vec x_m)\tilde\psi_j(\map(\vec x_m,
  \vec\omega_m))\nonumber\\
 &= \int_{\domainX} \expect[\psi_i(\vec x) \tilde\psi_j(\map(\vec
  x_m))\rho(\vec x)]\;d\vec x \nonumber\\
 & = \left\langle \psi_i, \expect[\tilde\psi_j\circ\map]\right\rangle_\rho,
\end{align}
where $\expect$ denotes the expected value over the stochasticity in the
dynamics, and represents the integral taken over the probability
space.

In this formulation, the connection to the Koopman operator appears in
(\ref{eq:A-limit}).
The Koopman operator  was originally defined for Hamiltonian
systems~\cite{Koopman1931,Koopman1932}, but in recent years has also
been applied to dissipative
systems~\cite{Mezic2013,Mezic2005,Rowley2009,Williams2014} and those
with stochastic dynamics~\cite{Mezic2005,Williams2014};
this latter formulation is most relevant here.
In this application, the Koopman operator, which we denote as
$\koopman$, is defined for a discrete-time {\em Markov process} with
the evolution operator $\map$.
The appeal of studying the Koopman operator instead of  $\map$, is
that $\koopman$ is linear even when $\map$ is nonlinear.
However, the Koopman operator acts on scalar observables, such as the
$\psi_k$ or $\tilde\psi_k$, which map state space to scalars, and  is infinite dimensional even when $\map$ is finite
dimensional.

For the observable $\tilde\psi:\domainY\to\mathbb{R}$, the action of the
Koopman operator is 
\begin{equation}
  \label{eq:koopman}
  \koopman\tilde\psi = \expect[\tilde\psi\circ \map],
\end{equation}
where $\expect$ is the expectation over the stochastic dynamics in
$\map$, and $\koopman{\tilde\psi}:\domainX\to\mathbb{R}$ is another function {\em defined
on a different domain}.
With the Koopman operator, the $ij$-th element of $\mat{A}$ can be
written succinctly as 
\begin{equation}
  \label{eq:Adef-koopman}
  \mat{A}^{(ij)} = \left\langle\psi_i, \koopman\tilde\psi_j\right\rangle_\rho,
\end{equation}
and due to the linearity of the Koopman operator, the objective
function can be written as 
\begin{equation}
  \label{eq:objective-koopman}
  g(\setX, \setY) = \left\langle \setX, \koopman\setY\right\rangle_\rho,
\end{equation}
which leads to the  optimization problem:
\begin{subequations} 
\label{eq:relaxed-koopman}
\begin{align}
  \max_{\setX, \setY} \quad& \left\langle \setX, \koopman\setY\right\rangle_\rho\\
  \text{subject to:}  \quad& \left\langle 1, \setX\right\rangle_\rho = \cutoffX, \\
                      & \left\langle 1, \setY\right\rangle_\nu = \cutoffY, \\
                      & \left\langle \setX, \setX \right\rangle_\rho =
                      \left\langle \setY, \setY\right\rangle_\nu = 1,
\end{align}
\end{subequations}
that we would  solve given an infinite amount of data and a
complete set of basis functions.

The benefit of this formulation is that it makes
the similarities between our method and the one presented in
Ref.~\citenum{Froyland2013} clear.
Because the Koopman operator, $\mathcal{K}$,  is the adjoint of the
(modified) Perron-Frobenius operator, $\mathcal{L}$, used there, the
objective function can either be written as $\left\langle \setX,
  \koopman\setY\right\rangle_\rho$ or $\left\langle
  \mathcal{L}\setX,\setY\right\rangle_\nu$.
The latter expression is of the same form as the objective function used by \citet{Froyland2013}, and could be equivalent provided the noise added to the system is chosen appropriately.

Furthermore, the problem in (\ref{eq:relaxed-discrete}) can also be
derived by approximating the Koopman operator using Extended Dynamic
Mode Decomposition~\cite{Williams2014}  with the set of snapshot pairs
$\{(\vec x_m, \vec y_m)\}_{m=1}^M$, the basis functions $\psi_k$ and
$\tilde\psi_k$, and using Monte-Carlo
integration to approximate any needed inner products. 
As a result and assuming $\setX$ and $\setY$ are smooth, our procedure will converge to the true solution at a rate of either $\mathcal{O}(\sqrt{M})$, if the error is dominated by errors in the integrals, or at a rate determined by set of basis functions used to approximate $\setX$ and $\setY$.
In principle, however, any method that can approximate the action of the Koopman
operator directly from data, such as Generalized Laplace
Analysis~\cite{Budisic2012}, could also be
used to compute coherent sets.
This would lead to a different optimization problem and a different rate of convergence.

\section{Example Applications}
\label{sec:examples}

In this section, we consider three examples that demonstrate
the efficacy of our method. 
The first is the double gyre, which is defined on a fixed domain, and
frequently used as a test problem for coherent set identification.
The purpose of this example is to demonstrate that the approach
described here produces coherent sets that are similar to the sets
produced by GAIO using an approximation of the Perron-Frobenius operator.
Next, we consider the Bickley Jet, which is an idealized but more realistic
problem where the data are not initially confined to some trapping
region, and therefore,  mesh-free approximations of $\setX$ and
$\setY$ becomes necessary.
Next, we consider the example of numerically simulated ``drifters''
in the Sulu Sea, which is a realistic example of how we envision this
technique being used in practice. 
In this example, our objective is to  identify an eddy
that is already known to exist in the time frame of the simulation.

\subsection{Choosing the Basis Functions}

As mentioned previously, one important facet of this procedure is the
choice of the  basis functions that are the building blocks for $\vec\psi_X$
and $\vec{\psi}_Y$. 
In each of these  problems, we use a basis set of thin-plate
splines, which are functions of the form:
\begin{subequations}
  \label{eq:thin-plate}
\begin{align}
  \psi_k(\vec x) &= r^2\log(r), \quad \text{ where } r = \|\vec x -
  \vec \xi_k\|, \\
  \tilde\psi_k(\vec y) &= \tilde r^2\log(\tilde r), \quad \text{ where
  } \tilde r = \|\vec y - \vec{\tilde \xi}_k\|,
\end{align}
\end{subequations}
where $\vec\xi_k$ is the $k$-th radial basis function (RBF) center, which is
a vector in $\mathbb{R}^2$ that defines the center-position of the thin
plate spline.
We also define the basis functions used to construct $\setY$ in a
similar manner, but call the associated centers $\vec{\tilde \xi}_k$.  

Thin plate splines are a special case of polyharmonic splines that are tailored for problems in $\mathbb{R}^2$, and commonly used for the interpolation of scattered data~\cite{iske2004multiresolution}.
Although they are not compactly supported, these functions have two useful properties: (i) they do not require the scaling parameter that many other radial basis functions do, and (ii) they do not require a computational mesh to be defined~\cite{Liu2010,Wendland1999,Fasshauer1996}.
The $\vec\xi_k$ and $\vec{\tilde\xi}_k$ are chosen by applying $k$-means
clustering~\cite{Bishop2006} to the collection of $\vec x_m$ and $\vec
y_m$ snapshots respectively.
$k$-means clustering partitions a set of data into $k$-sets, which are
chosen to minimize the total distance between the points and the mean
of the set they are  assigned to.
We use the set of means that result from this procedure as the $\vec
\xi_k$ and $\vec{\tilde{\xi}}_k$ respectively. 

To determine the number of basis functions, $K_X$ and $K_Y$, we first
choose a ``conservative'' pair of values, say, $K=K_X =K_Y = 5$.
Next we compute the leading singular value of the $\mat{\hat{A}}$
associated with the basis sets generated by this value of $K$.
If the leading singular value is one, then we increment $K$, and
repeat the process until this constraint fails to hold.
The results in this section are from the largest values of
$K_X$ and $K_Y$ that did not violate our sanity check, which as a rule of
thumb, corresponds to between 5-20 data points
per basis function. 
This procedure is {\em ad hoc}, but appears to produce a useful set of basis functions for the  examples presented in
this manuscript.

This is, of course, not the only possible choice of basis functions,
nor do we claim it is in any way optimal.
However, the benefit of using the basis elements associated with mesh-free methods, such as the thin
plate splines, is that they can be applied to problems on domains that
are not simple rectangles, which makes them suitable for a wide range
of applications.
Similar to GAIO, this allows us to apply the same procedure to all  of the examples that follow despite the fact only one of them is defined on a fixed domain.

\subsection{The Double Gyre}

\label{sec:double-gyre}

\begin{figure}[tb]
  \centering
  \includegraphics[width=0.7\columnwidth]{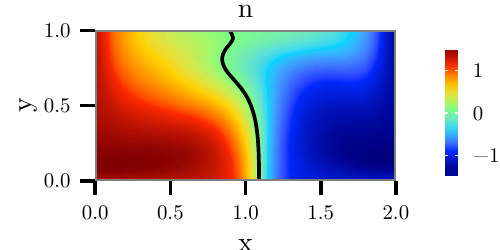}
  \caption{The function equivalent to $\setX$ computed using GAIO
    with  262,144 indicator functions (and the equivalent of 104,857,600 data pairs).
    The black line denotes the zero level set of both functions, which
    would be used to partition the domain into two coherent sets.
}
  \label{fig:gaio-double-gyre}
\end{figure}

\begin{figure*}[tb]
\includegraphics[width=0.8\textwidth]{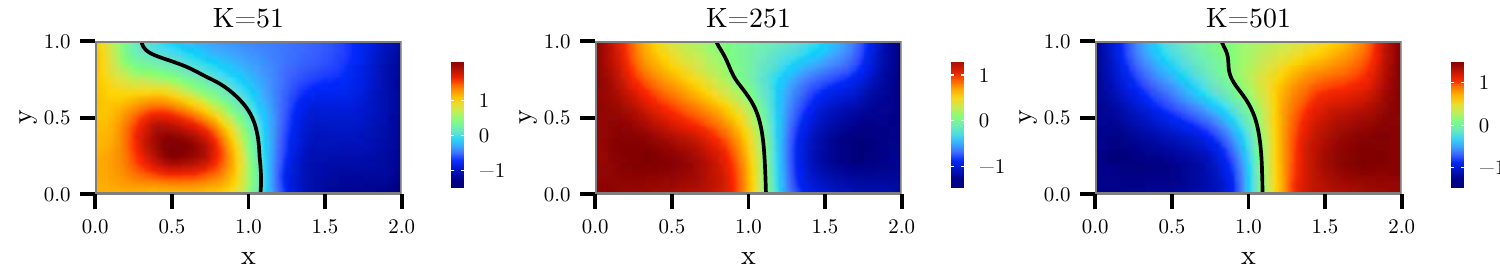} 
\caption{The function $\setX$ for the double gyre computed
  using $51$, $251$, and $501$ basis functions (e.g., 50 thin plate
  splines and the constant function) with $\cutoffX=\cutoffY=0$.
  The black line indicates the zero level set, which partitions the
  domain into the two sets $X_1$ and $X_2$.
  These results should be compared with the ``true'' solution in Fig.~\ref{fig:gaio-double-gyre}.
}
\label{fig:gyre-data} 
\end{figure*}

\begin{figure*}[tb]
\includegraphics[width=0.8\textwidth]{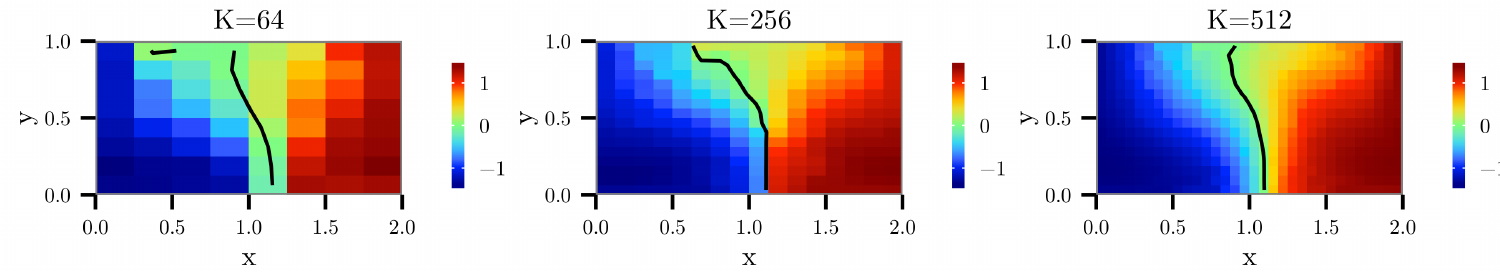} 
\caption{The function equivalent to $\setX$ 
  obtained using GAIO with $64$, $256$, and $512$ basis
  functions and 25 uniformly distributed data points per basis
  function; the images here are a benchmark for our results, which are
  shown in Fig.~\ref{fig:gyre-data}.
  }
\label{fig:gyre-gaio-data} 
\end{figure*}

Our first example is the double gyre, whose governing equations
are:
 \begin{subequations} 
\begin{align}
\dot{x} & =-\pi A\sin(\pi h(x,t))\cos(\pi y),\label{eq:gyre}\\
\dot{y} & =\pi A\cos(\pi h(x,t))\sin(\pi y)\frac{\partial h}{\partial x},
\end{align}
\end{subequations} where $h(x,t)=\epsilon\sin(\omega t)x^{2}+(1-2\epsilon\sin(\omega t))x$
with ${\epsilon=0.25}$, $\omega=2\pi$, and $A=0.25$. 
In these equations,  $x\in[0,2]$ and $y\in[0,1]$.
The double gyre
with these parameters is a frequently used test case for coherent
structure computations. See, for example, Refs.~\citenum{Ma2014,Bollt2013,Ma2013},
which compute coherent sets (albeit with slightly different definitions)
for this problem and parameters.

The purpose of this example is to demonstrate that the computational
procedure outlined in Sec.~\ref{sec:conceptual-pair} produces coherent
sets that are similar to those produced using the
definition in
Refs.~\citenum{Froyland2007,Froyland2010,Froyland2013,Bollt2013},
from a more limited amount of data.
For the purposes of comparison, Fig.~\ref{fig:gaio-double-gyre} shows
the  equivalent of $\setX$ identified by GAIO, which uses 262,144
indicator functions
and a total of 104,857,600 data pairs (i.e., 400 points per function). 
Because it uses indicator functions, a large basis set and,
hence, a large amount of data is required if the resulting functions are
to look smooth.
%,
By using tree-like data structures, this computation can be performed
quickly even with hundreds of millions of data points~\cite{Dellnitz2001}.
However, in applications where experimental rather than numerical data
is being used, obtaining such a large set may not be possible.

To  highlight the performance of the method, we apply it  with 51, 251, and 501 basis functions
(e.g., 50 thin plate splines and the constant function) using 1000,
5000, and 10,000 data pairs respectively.
The data at the initial time (i.e., the $\vec x_m$) are chosen by
randomly selecting initial conditions from a uniform distribution on
state space. 
Because the governing equations are discrete, we make the system
stochastic by adding noise all the $\vec x_m$ and $\vec y_m$.  
In this example, we make 20 copies of each of our data pairs and
perturb the data by adding a random vector chosen from a normal
distribution with a standard deviation of $10^{-3}$; with the basis
sets we will use, neither the number of copies nor the precise nature
of the noise will have a qualitative impact on the resulting functions. 
As a result, the values of $M$ in our computation are 20,000, 100,000,
and 200,000, which accounts for these additional copies, but similar
results could be obtained in the noiseless case with $M =1000$, 5000,
and 10,000.
Finally, we impose that $\cutoffX=\cutoffY =0$ in order to facilitate comparison with GAIO.

Figure~\ref{fig:gyre-data} shows the function, $\setX$,
obtained using the three sets of data listed above.
The black line denotes the $\setX  = 0$ level set,
which is used to partition state space into the pair of coherent sets.
As shown above, our approximations of $\setX$ appear to be converging to a particular
function as the number of basis functions and data points increases;
when run with 1,001 basis functions and 20,000 sets of data
($M=4\times 10^5$) the resulting $\setX$ is qualitatively similar to the function obtained with 501 basis functions. 

As a benchmark for our approach, Fig.~\ref{fig:gyre-gaio-data} shows
the function equivalent to $\setX$  computed using GAIO
with 64, 256, and 512 basis functions with 25 data points per basis
function initialized on a uniform grid.
This figure should be compared to Fig.~\ref{fig:gyre-data}, which used
slightly fewer basis functions (i.e., 51, 251, and 501) with 20
randomly distributed initial conditions per basis function.
As a result, each column in Fig.~\ref{fig:gyre-data} and
Fig.~\ref{fig:gyre-gaio-data} are comparable.
The most apparent difference between these two sets of results is the
smoothness of $\setX$, which are both clearly
discontinuous in Fig.~\ref{fig:gyre-gaio-data} due to the basis set
that is implicitly chosen by GAIO.
More importantly for experimental applications, we obtain a solution
that is {\em qualitatively} similar to the ``true'' solution with only 251
basis functions, while GAIO requires at least twice that amount.

We should note that there are quantitative differences between the coherent
sets identified using our method and the ones identified by GAIO.
In particular,
there are small quantitative differences in the zero level sets near
the point $x=1$ and $y=1$ for $\setX$ and $x=1$ and $y=0$ for
$\setY$. 
Part of this difference is due to the noise added
to the data; our approach explicitly adds normally distributed
perturbations, and GAIO {\em implicitly} adds noise that is related to
the width of each subdomain~\cite{Froyland2013}.
As a result, the part of the error due to differences in the added
noise would not vanish even if the amount of data was effectively
infinite.
However, there is also a difference in the spaces spanned by the thin
plate splines used here and the indicator functions used by GAIO; in practice, this difference has a larger impact on the resulting sets.
In this example, a large number of thin plate splines would be required
to capture the sudden ``bulge'' that occurs in $\setX$ near
the edge of the domain.

As shown here, our approach compares favorably to transfer operator
based methods for coherent set identification, and produces coherent
pairs that are qualitatively similar to the ones identified by those
methods.
Our approach uses a smaller number of globally supported basis functions, and appears to converge more rapidly because $\setX$ and $\setY$ are smooth in this problem.
As a result, useful and accurate approximations of these functions can be obtained with fewer data points than other methods may require.

\subsection{The Bickley Jet}

\label{sec:rossby}

\begin{figure*}[tb]
\centering \includegraphics[width=1\textwidth]{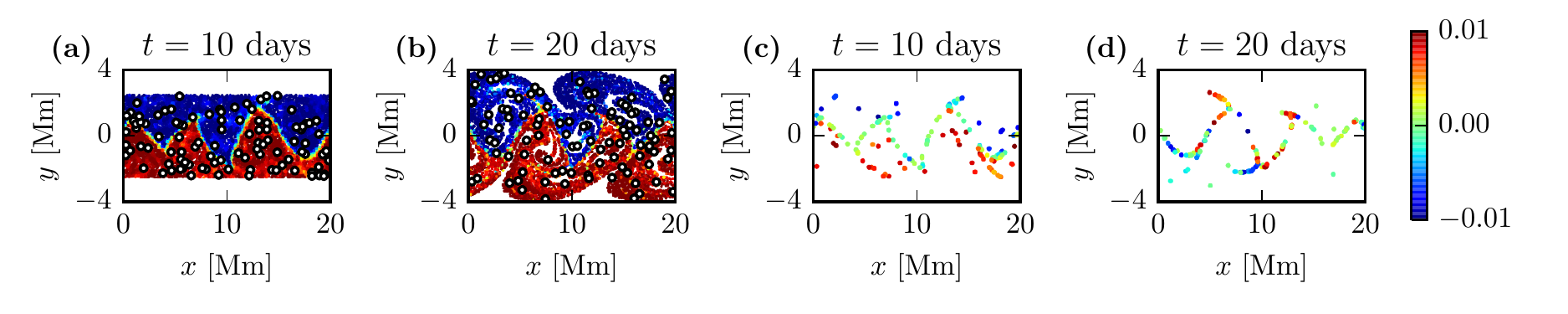} 
\caption{(a) The $10^{4}$ data points that comprise the data set colored by
the numerically computed approximation of $\setX$.
(b) The set of points in (a) at their new positions at $t=20$ days;
note that these points are still colored by $\setX$ rather than $\setY$.
(c) The set of 158 mis-classified points (i.e., those that ``leaked'' out over
the 10 day window) at $t = 10$ days.
(d) The same set of points at $t=20$ days.
The white circles in the left two images represent 100 of the 1000
$\vec\xi_k$ and $\vec{\tilde \xi}_k$ that were used as the centers of
the radial basis functions.
Despite the complexity of the ``true'' coherent sets and the relatively small
number of basis functions, there is little mixing between the
numerically computed sets, and most of the mixing occurs on the boundary between the
sets or in thin filaments extending into either side.
}
\label{fig:rossby} 
\end{figure*}

In this example, we demonstrate the effectiveness of this method by
computing a pair of coherent sets in the Bickley Jet flow which is a
dynamically-consistent approximation of an idealized stratospheric flow~\cite{Rypina2006}. 
We are concerned with sets that are optimal for the interval $t\in[10,
20]$ days, which was chosen so that these results may be compared with
pre-existing results~\cite{Bollt2013}.
This idealized system is Hamiltonian: 
\begin{subequations}
  \begin{align}
    \frac{\partial x}{\partial t}=-\frac{\partial\Phi}{\partial y},\label{eq:rossby}\\
    \frac{\partial y}{\partial t}=\frac{\partial\Phi}{\partial x},
  \end{align}
  where 
  \begin{align}
    \Phi(x,y,t) &=c_{3}y+U_{0}L\tanh(y/L) \\
    &+ A_{3}U_{0}L\sech^{2}(y/L)\cos(k_{3}x)\nonumber\\
    &+A_{2}U_{0}L\sech^{2}(y/L)\cos(k_{2}x-\sigma_{2}t)\nonumber\\
    &+A_{1}U_{0}L\sech^{2}(y/L)\cos(k_{1}x-\sigma_{1}t),\nonumber
  \end{align}
\end{subequations}
with $U_{0}=62.66$~m/s, $L=1770$~ km, $c_{2}=0.205U_{0}$,
$c_{3}=0.7U_{0}$, $A_{1}=0.075$, $A_{2}=0.4$, $A_{3}=0.2$, $k_{1}=2/r_{c}$,
$k_{2}=4/r_{c}$, $k_{3}=6/r_{c}$, $r_{c}=6.371$, $\sigma_{2}=k_{2}(c_{2}-c_{3})$,
and $\sigma_{1}=\frac{1+\sqrt{5}}{2}\sigma_{2}$. 
See \citet{Rypina2006} for an explanation of these parameter values. 
Our initial data are $10^{4}$ uniformly distributed on $x\in[0,20]$~Mm and
$y\in[-2.5,2.5]$~Mm at $t=10$ days, which we augment by making 20 copies where both $\vec x_m$ and $\vec y_m$ are perturbed randomly using numbers drawn from a normal distribution with a standard deviation of $10^{-3}$.
Even without noise, many initial conditions will leave this window, so $\domainX\neq \domainY$, and choosing a different set of basis functions to represent $\setX$ and $\setY$ is critical.
We use $10^{3}$ thin plate splines
whose centers are chosen using the $k$-means procedure outlined at the
start of this section; a subset of these locations are indicated by the white dots in Fig.~\ref{fig:rossby}, which makes it clear that the resulting distributions are qualitatively different at the two times.

Figure~\ref{fig:rossby} shows the results obtained with these basis
functions and data. 
For this problem, the geometry of the
coherent pair is more complex, and both sets have
a  ``sawtooth'' pattern.
Like before, the function in Fig.~\ref{fig:rossby}a changes
rapidly in value from approximately -0.015 to 0.015 at the boundary
between the coherent pair. 
Note that the color-scales in those images are restricted to -0.01 to 0.01 rather than the full range of values.
The sets identified by partitioning the data based on the sign of
$\setX$ with with $10^{4}$ data points compares favorably with those in
Ref.~\citenum{Bollt2013}, which uses over a million data points.
In particular, we partition the computational domain into the two
subdomains that are located above and below the oscillating
jet near $y=0$ that separates them.
 Furthermore, as shown in the figure, this approximation is ``good
enough'' that only 158 of the $10^4$  numerically classified
points ``leak'' out of the set they were assigned to;
similar to Refs.~\citenum{Froyland2010a,Bollt2013}, this leakage occurs
either on the boundary between sets or  on thin filaments that penetrate into either side.

The difference between the Bickley Jet and the double gyre example in Sec.~\ref{sec:double-gyre} is where the initial data were located.
For the double gyre, $\domainX$ was a trapping region, and $\domainX = \domainY$, but that is not the case in this problem.
While the definition of a coherent set in Sec.~\ref{sec:conceptual-pair}, one must also choose basis functions associated with ``mesh-free'' numerical methods if this computational procedure is to be viable. 
In this example, we used thin-plate splines, and recovered a good approximation of the sets obtained via GAIO-like methods with far fewer data points.

\subsection{Numerical Drifters in the Sulu Sea}
\label{sec:phil-sea}

\begin{figure*}[t]
\includegraphics[width=0.8\textwidth]{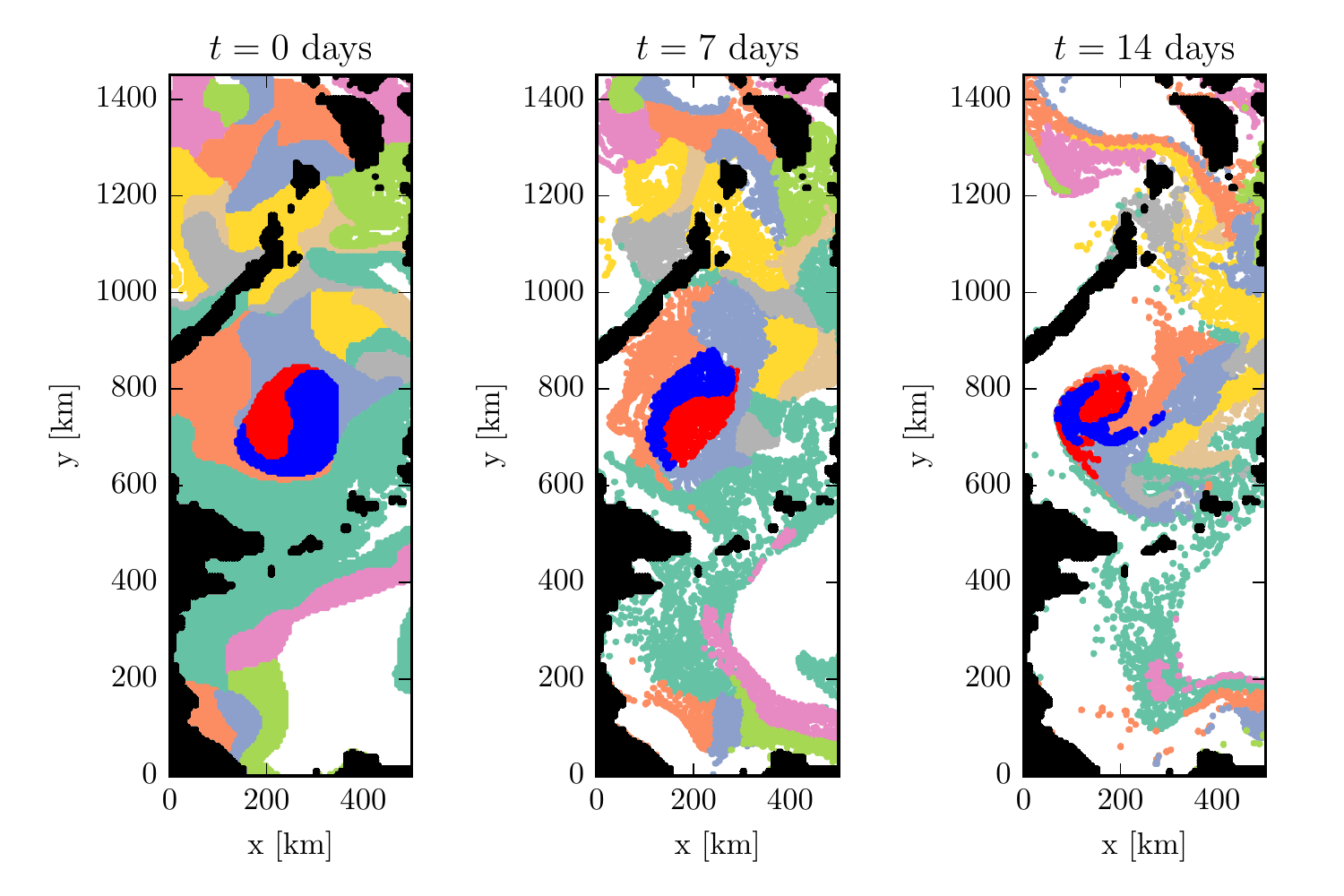} 
\caption{
The three images above show the evolution of the ``complete'' set of 25,146 data points at $t=0$, 7, and 14 days.
Two of the 27 coherent sets identified by the method, which correspond
to the eddy of interest, are shown in red and blue; the other 25 sets, which may or may not have a physical interpretation,
are indicated by the other colors.
In all three images, the black points denote land.
It should be noted that these coherent sets are only optimal from $t=0$ to 7 days.
At $t=14$ days, this lack of optimality can be seen in the long filaments that have formed in the blue and red sets.
}
\label{fig:phil-full} 
\end{figure*}

In this section, we consider a more realistic example generated by a numerical model ROMS\cite{Shchepetkin2005} for the Philippine Archipelago\cite{Rypina2010}.
Similar to the work of~\citet{Rypina2010}, the objective here is to
use our coherent set definition to identify a mesoscale anticyclonic
eddy that was present in the Sulu Sea.
Our data come in the form of numerically simulated drifters, that
are sampled once every week.
These drifters are randomly and uniformly distributed over the computational
domain.
In an experiment rather than a simulation, initializing thousands of drifters
is infeasible, and so the data available are truly limited in quantity.
As such, we consider two cases: the first consists of 25,146 tracers
randomly but uniformly distributed over the computational domain,
and constitutes a ``data rich'' example, which we will use to determine
the ``true'' coherent sets. 
Then we will reduce the amount of data
to 400 uniformly but randomly distributed initial conditions, which
is a more realistic  amount of data, and compare the
results obtained from this ``data poor'' set with the full data.
As before, we make 20 copies of the data, and add normally distributed
noise with a standard deviation of 100 m to both the $\vec x_m$ and the $\vec y_m$.

\begin{figure*}[t]
\includegraphics[width=0.8\textwidth]{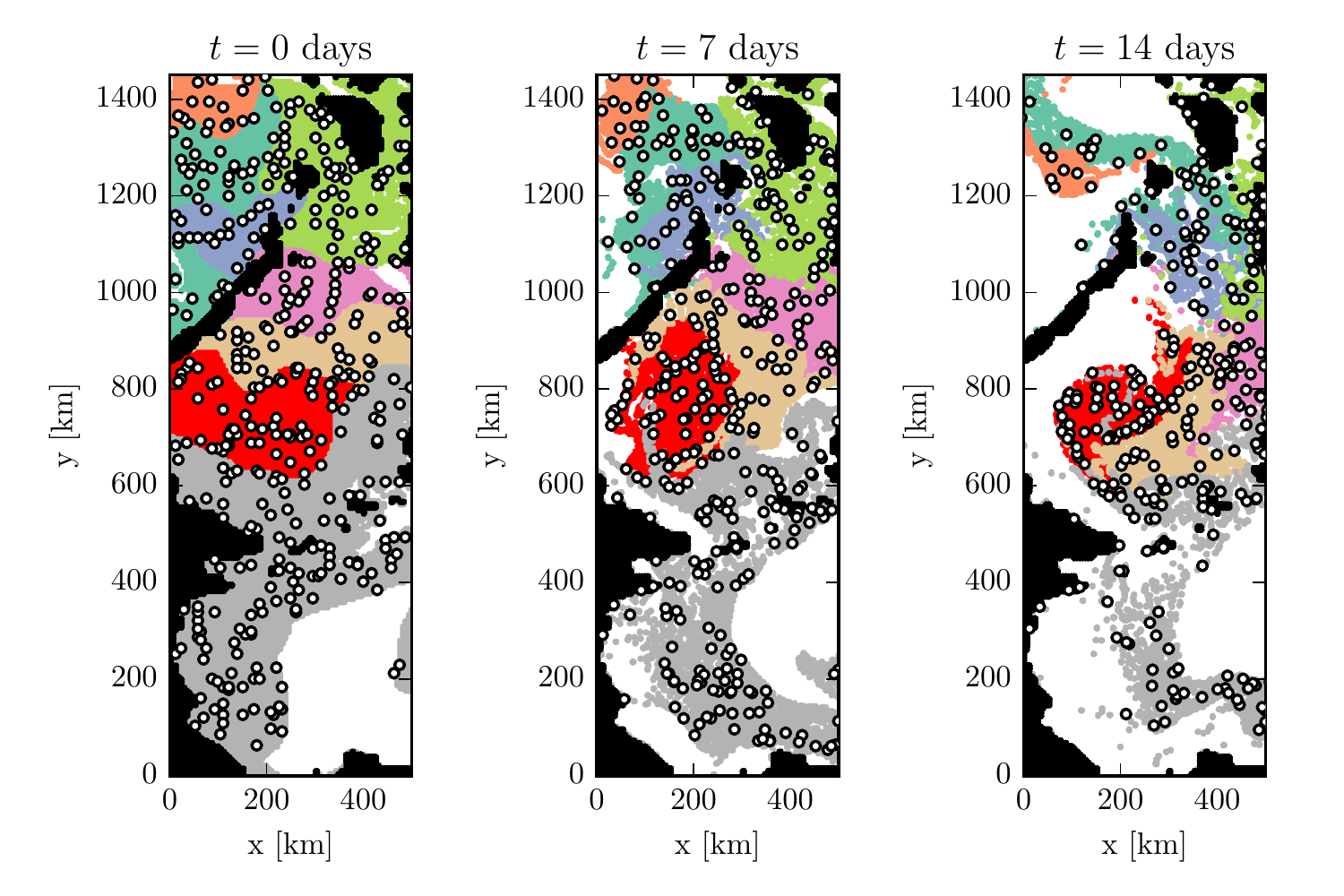}
\caption{This is a reproduction of Fig.~\ref{fig:phil-full} using only
  the 400 data points indicated by the white dots.
  As in that figure, the other 24,746 points are colored based 
  on which coherent set they are assigned to by the approximation 
  of $\setX$ obtained from the indicated set of 400 points.
  Note that the eddy is once again identified, but is now contained
  within a single coherent set that is shown in red.
 }
\label{fig:phil-reduced} 
\end{figure*}

In this example, we are interested in identifying a mesoscale eddy with a 100-km radius within a much larger, 500 km by 1500 km,  domain.
While both the ``rich'' and ``poor'' data sets can
be partitioned into a pair of coherent sets,  due to the implicit constraint on the size of
these sets, neither will immediately identify the eddy of interest.
Therefore, it becomes necessary to iterate the procedure and to further subdivide  space until
the size of the coherent sets is on the same order as the eddy. 
As a result, we will iterate up to four times using the procedure
described in Sec.~\ref{sec:conceptual-pair} and  Refs.~\citenum{Bollt2013,Ma2013}.
For each of the iterates, we choose $\cutoffX$ and $\cutoffY$
to maximize the fraction of consistently classified data pairs.
We limit the range of values that the $\cutoffX$ and $\cutoffY$ can
take on so that the resulting sets contain (roughly) the same number
of points (i.e., the  smaller set must contain at least 25\% of the total data). 
This additional restriction is {\em ad hoc}, and meant to
prevent the algorithm from selecting ``trivial'' sets with only
a handful of isolated data points.
 Initially we use 250 basis functions
for the full data, and 120 for the reduced data set of 400 points.
After every subsequent iteration, we divide the number of basis functions
used in the computation by two and require that the number of basis
functions is no more than 30\%  of the data points; as a result,
later iterations are performed on smaller domains and with fewer basis functions.

In Fig.~\ref{fig:phil-full}, we show the hierarchy of coherent sets
that is optimal for the $t=0$ to $t=7$ days time window using the full
set of the 25,146 data points available, where each of the points are
``colored'' by which of the 27 sets they were assigned to.
Note that the number of sets is a result of the recursion procedure outlined above; more or fewer sets can be generated by changing the amount of allowable ``leakage,'' the maximum number of recursive iterations, and the cutoff points $\cutoffX$ and $\cutoffY$. 
There is an additional plot of the data at $t=14$ days that demonstrates that the coherent sets identified
by our method remain coherent even at longer times.
We should reiterate, however, that these sets are, by construction,
only optimal from $t=0$ to 7 days; the results at $t=14$ days are
extrapolation, and not guaranteed to still be coherent at that time.
In this example, the persistence of the identified coherent sets until
day 14 is consistent with \citet{Rypina2010}  who also found the eddy to be present over a 2-week period.
Note that even at $t=0$ the data has ``holes,'' which are due either 
to the presence of land, which is indicated by the black regions, or 
because initial conditions at those regions leave the computational
window in Fig.~\ref{fig:phil-full} before a week has elapsed.

The identified coherent sets are ``optimal''
sets, but that does not necessarily mean that all of them necessarily have a simple and straightforward physical interpretation.
However, one physically meaningful pair of coherent sets is indicated 
by the red and blue regions  near  $x=200$ km and $y=700$ km in the
figure, and corresponds to the eddy identified by \citet{Rypina2010}.
From $t=0$ to $t=7$, the red and blue sets move
counterclockwise around each other without much stretching and
folding, which would be typical for a cyclonic motion associated with
an eddy. 
Note however, that because the sets are not optimal at $t=14$
days, they begin to leak out of the eddy at that time forming long filaments. 
To summarize, with a large amount of drifter data, such as the 
dataset generated numerically using the velocity field, this 
approach can produce a partition of state space that contains physically meaningful time-varying sets.

However, the purpose of this manuscript is to demonstrate that these
results can be obtained with limited quantities of Lagrangian data, so
we repeat the computation above with 400 data points instead of the
full set of 25,146.
We follow the same iteration procedure as before, and compute
additional sets by recursing up to four times provided the identified coherent
sets ``mis-classified'' at most 5\% of the data points available to each stage of the recursion procedure (e.g., the first level of the full 25,146 point data set was allowed to mis-classify up to 1,257 points, but the first level of the 400 point data set is only allowed to mis-classify 20).
The results of this computation are shown in
Fig.~\ref{fig:phil-reduced}.
To aid the eye, the colored points are, once again, the complete data
set that is shown in Fig.~\ref{fig:phil-full}, where the colors
denote the various coherent sets.
The 400 points used in the computation are indicated by the large
white dots in the figure.

As shown, the relatively small amount of data and the concomitant
reduction in the number of basis functions has had an impact on the
resolution and accuracy of the resulting method. 
Visually, the coherent sets we identify are larger in area than
those that we obtain with full data because fewer iterations of the
coherent set algorithm can be performed before our sets allow more
than 5\% of their points to escape. 
Nonetheless, we once again identify the eddy, which is
now indicated solely by the red set.
Once again, this set is only optimal from $t=0$ to 7 days, and long
filaments are again visible at $t=14$ days.

It should be noted that whether the eddy is contained within a
single coherent set or a pair is determined by the
basis functions and data provided to the method.
In the ``data rich'' example above, this subdivision occurs at the final step of the procedure, so the red and blue sets in Fig.~\ref{fig:phil-full} can be merged by terminating the procedure one iteration sooner.
However changes in computational parameters such as the location or number of the thin plate splines can result in this subdivision occurring before the last iterate.
Due to the recursive nature of the procedure, the data are assigned to coherent sets in a ``greedy'' manner, and once the eddy has been subdivided, it will remain so in all future iterates.
Other procedures for identifying multiple sets that do not have this limitation have been developed~\cite{deuflhard2000identification,deuflhard2005robust}, but their integration within the framework presented here will be the focus of future work.

In this section, we considered a more realistic example: numerically simulated drifters in the Sulu sea.  
First, we applied our procedure to a relatively large set of data, and demonstrated that, under ideal conditions, it was able to identify an eddy that is known to exist in this flow.
Next, we limited the amount of data to 400 randomly chosen drifters.
The cost of working in this ``data poor'' regime is a loss of resolution; fewer coherent sets could be identified before the amount of ``leakage'' grew past our threshold.
In the end, however, our procedure once again identified the eddy of interest.

\section{Conclusions}

\label{sec:conclusions}
In this manuscript, we presented a method for computing
coherent sets that are optimal over a finite interval in time, which is conceptually related to that of \citet{Froyland2013}.
However, our interest is in the ``data poor'' regime, which is
common in problems involving experimental, rather than computational,
experiments.
In the double gyre example, we demonstrated that the coherent sets
identified using a limited number of thin plate splines agreed well
with the sets obtained using a larger number of indicator functions.
The benefit of using thin plate splines or other mesh-free basis
functions is that they can also be used in problems where a
computational grid is not easily defined.
This is useful in the second example involving the Bickley Jet, where
the initial domain is not a trapping region, and the domain at the
final time resembles a ``sawtooth.''
By using radial basis functions, the same procedure used for the
double gyre can also be used here without alteration.
Our final example is identifying an eddy in the Sulu Sea, which
possessed a changing computational domain in combination with a
relatively small coherent set of physical interest. 
In that example, we also demonstrated that our approach can
identify the eddy of interest even with relatively small amounts of
data that approach the number of drifters used in recent massive drifter deployment experiments~\cite{Rypina2014eulerian,Poje2014submesoscale}.

In all three examples, the noise added to the $\vec x_m$ and $\vec y_m$ was normally
distributed with a standard deviation that was small compared to the
spatial scales on which the problem was defined. 
However, it appears that the ``noise'' in the dynamics introduced implicitly by our basis functions typically has a far larger impact on the resulting coherent sets.
If this is not the case, one improvement to the procedure would be to use observation-based spatially-dependent anisotropic diffusivities (see, for example,
\citet{Rypina2012eddy}) to represent the stochastic portion of the
flow rather than arbitrarily choosing a distribution as we do here.

Because of the crucial role transport barriers play in understanding
systems with chaotic mixing, algorithmic methods for identifying these
barriers are useful tools for researchers in application areas like
geophysical fluid dynamics, combustion, and even those focused on
ecological problems.
In some situations, one either knows or can approximate the velocity
field of the flow, which enables standard techniques
and software packages such as FTLE fields or GAIO to be used. 
However, in other applications, the velocity field cannot be obtained
analytically or numerically, and Lagrangian data from drifting buoys
are all that is available to us.
Ultimately, algorithms such as the one presented here are the first
steps  towards adapting the techniques we would use in a data rich environment  for use in practical problems where the needed Lagrangian data
are  sparse and difficult/expensive to obtain.

\section*{Acknowledgments}
The authors would like to acknowledge I.G. Kevrekidis for helpful
discussions on set-oriented methods and suggestions for this manuscript.
M.O.W. gratefully acknowledges support from the NSF (DMS-1204783).  
I.R. was supported by ONR (MURI award N000141110087), and by the NSF (grant 85464100).
C.W.R. was supported by AFOSR (grant FA9550-14-1-0289).

\appendix
\section{The Leading Singular Vectors of the Approximation}
\label{app:block-structure}

To prove that the first unit vectors are also singular vectors of
$\mat{\hat{A}}$, we must first prove that $\mat{\hat{A}}$ is block diagonal.
To show this structure arises, it is convenient to define the data matrices
\begin{equation}
\mat{\Psi_X} = 
\begin{bmatrix}
\vec\psi_X(\vec x_1)^T \\
\vec\psi_X(\vec x_2)^T \\
\vdots \\
\vec\psi_X(\vec x_M)^T
\end{bmatrix},
\quad 
\mat{\Psi_Y} = 
\begin{bmatrix}
\vec\psi_Y(\vec y_1)^T \\
\vec\psi_Y(\vec y_2)^T \\
\vdots \\
\vec\psi_Y(\vec y_M)^T
\end{bmatrix},
\end{equation}
where $\mat{\Psi}_X\in\mathbb{R}^{M\times K_X}$ and
$\mat{\Psi}_Y\in\mathbb{R}^{M\times K_Y}$.
Note that $\mat{A} = \frac{1}{M}\mat{\Psi_X}^T\mat{\Psi_Y}$, $\mat{G_X} =
\frac{1}{M}\mat{\Psi_X}^T\mat{\Psi_X}$, and $\mat{G_Y} =
\frac{1}{M}\mat{\Psi_Y}^T\mat{\Psi_Y}$.

Because the Cholesky Decomposition is unique, we can also write 
\begin{equation}
\mat{\Psi_X} = \sqrt{M}\mat{Q_X}\mat{L_X}^T, \quad \text{ and }\quad
\mat{\Psi_Y} =\sqrt{M} \mat{Q_Y}\mat{L_Y}^T,
\end{equation}
where $\mat{Q_X}$ and $\mat{Q_Y}$ are orthonormal matrices.  
While $\mat{Q_X}$ and $\mat{Q_Y}$, in general, differ, this is a QR
decomposition, so the first columns of $\mat{Q_X}$ and $\mat{Q_Y}$ are
normalized version of the first columns of $\mat{\Psi_X}$ and $\mat{\Psi_Y}$.
However, because of our choice of basis functions (in particular, $\psi_1 =
\tilde{\psi}_1 = 1$), the first columns of $\mat{\Psi_X}$ and $\mat{\Psi_Y}$ are
identical. 
Therefore, the first columns of $\mat{Q_X}$ and $\mat{Q}_Y$, which we refer to
as $\vec{q_X}^{(1)}$ and $\vec{q_Y}^{(1)}$, are also identical. 
Furthermore, $\mat{Q_Y}^T\vec{q_X}^{(1)}= \vec{\tilde e}_1$ and
$\mat{Q_X}^T\vec{q_Y}^{(1)}= \vec{e}_1$ by the orthonormality of $\mat{Q_X}$ and
$\mat{Q_Y}$. 
Finally, $\mat{\hat{A}} = \mat{L_X}^{-1}\mat{A}\mat{L_Y}^{-T} =
\mat{L_X}^{-1}\mat{L_X}(\mat{Q_X}^T\mat{Q_Y})\mat{L_Y}^T\mat{L_Y}^{-T}=\mat{Q_X}^T\mat{Q_Y}$.
Because the first columns of $\mat{Q_X}$ and $\mat{Q_Y}$ are identical 
\begin{equation}
\mat{\hat{A}} = \mat{Q_X}^T\mat{Q_Y} = 
\begin{bmatrix}
1 & \vec 0 \\
\vec 0 & \mat{\hat{A}}_{22}
\end{bmatrix},
\end{equation}
which shows the matrix $\mat{\hat{A}}$ possess the desired block structure.
Because of this structure, it is now clear that $\vec e_1$ and
$\vec{\tilde e}_1$ are singular vectors associated with the singular value
$\sigma = 1$.

\bibliography{coherent}
 
\end{document}